\documentclass[aps,prb,twocolumn,groupedaddress,floatfix,showpacs]{revtex4-2}
\usepackage[T1]{fontenc}
\usepackage{graphicx}
\usepackage{amsmath}
\usepackage{color}

\begin{document}

% Use the \preprint command to place your local institutional report
% number in the upper righthand corner of the title page in preprint mode.
% Multiple \preprint commands are allowed.
% Use the 'preprintnumbers' class option to override journal defaults
% to display numbers if necessary
%\preprint{}

%Title of paper
\title{Fractional Derivative Modification of a Drude Model}

% repeat the \author .. \affiliation  etc. as needed
% \email, \thanks, \homepage, \altaffiliation all apply to the current
% author. Explanatory text should go in the []'s, actual e-mail
% address or url should go in the {}'s for \email and \homepage.
% Please use the appropriate macro foreach each type of information

% \affiliation command applies to all authors since the last
% \affiliation command. The \affiliation command should follow the
% other information
% \affiliation can be followed by \email, \homepage, \thanks as well.
\author{Karol Karpi\'{n}ski}
\author{Sylwia
Zieli\'{n}ska-Raczy\'{n}ska}
\author{David Ziemkiewicz}
\email{david.ziemkiewicz@utp.edu.pl}

%\email[]{Your e-mail address}
%\homepage[]{Your web page}
%\thanks{}
%\altaffiliation{}
 \affiliation{Institute of
Mathematics and Physics, UTP University of Science and Technology,
\\Aleje Prof. S. Kaliskiego 7, 85-789 Bydgoszcz, Poland.}

%Collaboration name if desired (requires use of superscriptaddress
%option in \documentclass). \noaffiliation is required (may also be
%used with the \author command).
%\collaboration can be followed by \email, \homepage, \thanks as well.
%\collaboration{}
%\noaffiliation

\definecolor{green}{rgb}{0,0.8,0}

\begin{abstract}
A modification of the Drude dispersive model based on fractional time derivative is presented. The dielectric susceptibility is calculated analytically and simulated numerically, showing a good agreement between theoretical description and numerical results. The absorption coefficient and wave vector - key parameters describing the propagation of waves in such a medium - are shown to follow a power law in the frequency domain, which is a common phenomenon in many real life applications. The introduction of two separate parameters provides a more flexible model than some other approaches found in literature and is well suited for numerical implementation.
\end{abstract}
\maketitle

\section{Introduction}
Factional calculus - branch of mathematics devoted to study non-integer order derivatives and integrals, becomes an increasingly popular tool for analysis of various problems in physics, ranging from quantum mechanics and cosmology \cite{Herrmann2018} to electric circuits \cite{Liang2017} and electromagnetic wave propagation \cite{Tarasov}. An extensive review of recent developments is presented in \cite{Machado2011}. 

The application of fractional derivative in theoretical \cite{Ray2016} and numerical \cite{Padovan1987} description of mechanical waves has been extensively studied, including the propagation of acoustic waves \cite{Holm2011}. One of the key motives behind the development of fractional models is the observation that wave attenuation in many systems follows a power law with a non-integer exponent \cite{Chen2004,Song2018}, which cannot be described with standard time domain partial differential equations. Furthermore, many physical media exhibit hereditary features \cite{Rossikin1997}, where some physical property is dependent on the history of its previous values; fractional derivative provides a tool to analyse such systems that carry information about their present as well as past states \cite{Nigmatullin1992,Enelund1999,Gomez2018}.

Motivated by these developments, we propose an extension of the Drude model, which is one of the basic tools for describing the electric permittivity of metals. To better fit the experimental observations in complex media, several extensions of Drude model have been proposed, ranging from introduction of frequency-dependent parameters \cite{Gantzler2018} to the use of fractional derivative \cite{Guia2016}. The approach presented here introduces two fractional derivatives to the equations of motion describing the medium, resulting in a more general and flexible model. 
Moreover, the implementation of the medium description in Finite-Difference Time-Domain (FDTD) scheme \cite{Yee} is presented. The accuracy of simulation results is discussed.

\section{Drude model}
We assume that the medium (usually metal) contains some concentration $n$ of free charges $e$, with effective mass $m$. The equation describing their motion under the influence of external field $E$ is
\begin{equation}
m\ddot{\vec{r}}+\Gamma\dot{\vec{r}}=e\vec{E},
\end{equation}
where $\vec r$ is the charge position and $\Gamma$ describes dissipative processes. By introducing the polarization vector $\vec{P}=ne\vec{r}$, damping constant $\gamma=\Gamma/m$ and plasma frequency $\omega_p^2=ne^2/m$, one obtains
\begin{equation}\label{glowne}
\ddot{P}+\gamma\dot{P}=\omega_p^2E
\end{equation}
which, in the time domain, for harmonic wave $E=E_0e^{-i\omega t}$, $P=P_0e^{-i\omega t}$ yields
\begin{equation}
P_0 = \chi(\omega) E_0 = \frac{-\omega_p^2}{\omega^2 + i\gamma\omega}E_0,
\end{equation}
which is a standard expression for modelling electric susceptibility of metals \cite{Gantzler2018}. In our modified model, we introduce the fractional derivative operator $\mathbf{D}$ to the Eq. (\ref{glowne}) in all places where time derivative is used. In particular, we transform the Eq. (\ref{glowne}) into the form
\begin{equation}\label{glowne_m}
\gamma_\alpha\mathbf{D}^{\alpha+1} P+\gamma_\beta \mathbf{D}^\beta P=\omega_p^2E
\end{equation}
where $\mathbf{D}^\beta$ denotes fractional derivative of the order $\beta$ and $\alpha$, $\beta$ are real parameters. Standard relation is obtained for $\gamma_\alpha=\alpha=\beta=1$. In the frequency domain, the above relation leads to the susceptibility
\begin{equation}\label{koncowe_m}
\chi(\omega)=\frac{P_0}{E_0}=\frac{\omega_p^2}{\gamma_\alpha(-i\omega)^{\alpha+1} + \gamma_\beta (-i\omega)^\beta}.
\end{equation}
Note that the constants $\gamma_\alpha$, $\gamma_\beta$ ensure the proper dimensionality of the equation, e. g. \hbox{$\left[\gamma_\alpha\omega^{\alpha+1}\right]=\left[\omega^2\right]$} and \hbox{$\left[\gamma_\beta\omega^{\beta}\right]=\left[\omega^2\right]$}, so that $\chi$ remains a dimensionless quantity.

In contrast to other extensions of Drude model, where dissipation constant and/or plasma frequency are frequency-dependent quantities \cite{Gantzler2018}, we introduce additional frequency dependence by directly modifying the exponent of the frequency.

\section{Numerical implementation}
The Finite-Difference Time-Domain method (FDTD) consists of dividing the simulation space into grid points and calculating the values of electric and magnetic field at those points with evolution equations derived directly from Maxwell's equations, with some finite time step $\Delta t$ \cite{Yee}. Due to the fact that the algorithm is based on first principles, has a well-known sources of numerical errors and is easy to implement in parallel computing, it is one of the leading tools for analysis of complex optical and plasmonic systems.   
One of the methods to include complex media in FDTD simulations is ADE (Axillary Differential Equations) approach \cite{Alsunaidi}; the medium polarization $P$ is computed by numerically solving a partial differential equation describing its evolution in time. For regular Drude model, one can derive the evolution equation for $P$ from Eq. (\ref{glowne}). In the FDTD scheme, one has a set of discrete values of polarization $P_t$, $t=1,2,3...$. To obtain the first-order approximation of the first time derivative, one can use the relations
\begin{eqnarray}
\dot{P}_{t+1/2} \approx \frac{P_{t+1} - P_t}{\Delta t},\nonumber\\
\dot{P}_{t-1/2} \approx \frac{P_{t} - P_{t-1}}{\Delta t},\nonumber\\
\dot{P_t} \approx \frac{1}{2}(\dot{P}_{t+1/2} + \dot{P}_{t-1/2}).\label{avg_1}
\end{eqnarray} 
In a similar manner, one can define the second derivative as
\begin{equation}
\ddot{P_t} \approx \frac{1}{\Delta t}(\dot{P}_{t+1/2} - \dot{P}_{t-1/2}).\label{avg_2}
\end{equation}

In our modified approach, one has to first define the fractional derivative operator; from the standpoint of computation with discrete time step $\Delta t$, the most convenient definition is the truncated Grunwald-Letnikov derivative \cite{Garappa2019}
\begin{equation}
\mathbf{D}^\alpha f(t) = \lim_{\Delta t\rightarrow 0}\frac{1}{\Delta t^\alpha}\sum\limits_{k=0}^N (-1)^k\frac{\Gamma(\alpha+1)f(t-k\Delta t)}{\Gamma(k+1)\Gamma(\alpha-k+1)}
\end{equation}
where in numerical implementation $\Delta t$ is set to some finite value and $N$ is a suitably large number such that the components of the sum are negligibly small. Note that the left-sided derivative is used, where only previous values of $f(t)$ are needed. In other words, the model is casual. In terms of discrete values of $P$, one obtains
\begin{eqnarray}\label{disc_der}
&&\mathbf{D}^\alpha P_{t-\frac{1}{2}} = \frac{1}{\Delta t^\alpha} \sum\limits_{k=0}^N(-1)^k\frac{\Gamma(\alpha+1)}{\Gamma(k+1)\Gamma(\alpha-k+1)}P_{t-k}\nonumber\\
&& = \frac{1}{\Delta t^\alpha} \sum\limits_{k=0}^N\alpha_kP_{t-k}.
\end{eqnarray} 
By applying the above definition and the averaging procedure in Eqs. (\ref{avg_1}) and (\ref{avg_2}) to the Eq. (\ref{glowne_m}), one obtains the relations for derivatives
\begin{eqnarray}
&&\mathbf{D}^{\alpha+1} P_t = \frac{1}{\Delta t\Delta t^\alpha} \sum\limits_{k=0}^N\alpha_kP_{t-k+1}-\alpha_kP_{t-k}\nonumber\\
&&\mathbf{D}^\beta P_t = \frac{1}{2\Delta t^\beta} \sum\limits_{k=0}^N\beta_kP_{t-k+1}+\beta_kP_{t-k}
\end{eqnarray} 
and the equation of motion
\begin{eqnarray}
&&\frac{\gamma_\alpha}{\Delta t\Delta t^\alpha} \sum\limits_{k=0}^N\alpha_k(P_{t-k+1}-P_{t-k}) \nonumber\\
&&+\frac{\gamma_\beta}{2\Delta t^\beta}\sum\limits_{k=0}^N\beta_k(P_{t-k}+P_{t-k+1}) - \omega_p^2 E_t = 0.\qquad
\end{eqnarray}
By expanding and rearranging the terms, one obtains the evolution relation
%\begin{eqnarray}
%&&\frac{\gamma_\alpha}{\Delta t\Delta t^\alpha}\left(\alpha_0 P_{t+1} - \alpha_0 P_t + \alpha_1 P_t - \alpha_1 P_{t-1} + \sum\limits_{k=2}^N\alpha_k(P_{t-k+1}-P_{t-k})\right)\nonumber\\
%&& + \frac{\gamma_\beta}{2\Delta t^\beta}\left(\beta_0 P _{t+1} + \beta_0 P_t + \beta_1 P_t + \beta_1 P_{t-1} + \sum\limits_{k=2}^N\beta_k(P_{t-k+1}+P_{t-k}) \right)\nonumber\\
%&&  - \omega_p^2 E_t = 0
%\end{eqnarray}
%and rearranging the terms
%\begin{eqnarray}
%&&\left(\frac{\gamma_\alpha\alpha_0}{\Delta t\Delta t^\alpha}+\frac{\gamma_\beta\beta_0}{2\Delta t^\beta}\right)P_{t+1} + \left(\frac{\gamma_\alpha(\alpha_1-\alpha_0)}{\Delta t\Delta t^\alpha}+%\frac{\gamma_\beta(\beta_1+\beta_0)}{2\Delta t^\beta}\right)P_t+\left(\frac{-\gamma_\alpha\alpha_1}{\Delta t \Delta t^\alpha}+\frac{\gamma_\beta\beta_1}{2\Delta t^\beta}\right)P_{t-1} \nonumber\\
%&&+ \frac{\gamma_\alpha}{\Delta t\Delta t^\alpha}\sum\limits_{k=2}^N \alpha_k(P_{t-k+1} - P_{t-k}) + \frac{\gamma_\beta}{2\Delta t^\beta}\sum\limits_{k=2}^N\beta_k(P_{t-k}+P_{t-k+1}) = \omega_p^2 %E_t.
%\end{eqnarray}
%Finally, one obtains the evolution relation
\begin{eqnarray}
P_{t+1}=-\frac{\frac{\gamma_\alpha(\alpha_1-\alpha_0)}{\Delta t\Delta t^\alpha}+\frac{\gamma_\beta(\beta_0+\beta_1)}{2\Delta t^\beta}}{\frac{\gamma_\alpha\alpha_0}{\Delta t\Delta t^\alpha}+\frac{\gamma_\beta\beta_0}{2\Delta t^\beta}}&P_t&\nonumber\\ 
-\frac{\frac{-\gamma_\alpha\alpha_1}{\Delta t\Delta t^\alpha}+\frac{\gamma_\beta\beta_1}{2\Delta t^\beta}}{\frac{\gamma_\alpha\alpha_0}{\Delta t \Delta t^\alpha}+\frac{\gamma_\beta\beta_0}{2\Delta t^\beta}}&P_{t-1}&\nonumber\\
-\frac{\frac{\gamma_\alpha}{\Delta t\Delta t^\alpha}}{\frac{\gamma_\alpha\alpha_0}{\Delta t\Delta t^\alpha}+\frac{\gamma_\beta\beta_0}{2\Delta t^\beta}}\sum\limits_{k=2}^N\alpha_k &(P_{t-k+1} - P_{t-k})&\nonumber\\
-\frac{\frac{\gamma_\beta}{2\Delta t^\beta}}{\frac{\gamma_\alpha\alpha_0}{\Delta t\Delta t^\alpha}+\frac{\gamma_\beta\beta_0}{2\Delta t^\beta}}\sum\limits_{k=2}^N\beta_k &(P_{t-k+1} + P_{t-k})& \nonumber\\
+\frac{\omega_p^2}{\frac{\gamma_\alpha\alpha_0}{\Delta t\Delta t^\alpha}+\frac{\gamma_\beta\beta_0}{2\Delta t^\beta}} &E_t&.\label{koncowe_num}
\end{eqnarray}
which relates the new value of $P_{t+1}$ to the current value of $P_t$, the electric field $E_t$ and the history $P_{t-k}$. For the case of $\alpha=\beta=1$, $\alpha_k=\beta_k=0$ for $k>1$ and the relation simplifies to the standard form presented in \cite{Alsunaidi}.

\section{Results}
\subsection{Calculation of susceptibility}
\begin{figure*}[t!]
\centering
a)\includegraphics[width=.45\linewidth]{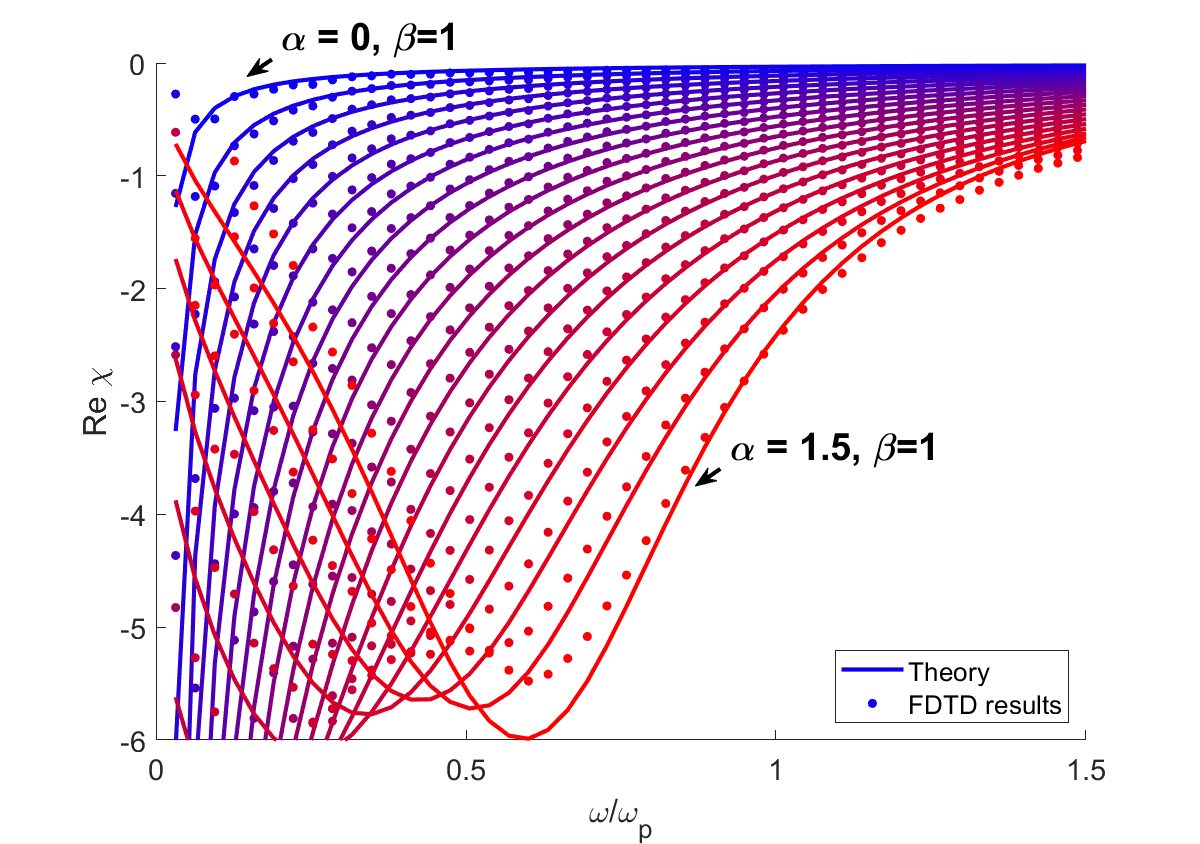}
b)\includegraphics[width=.45\linewidth]{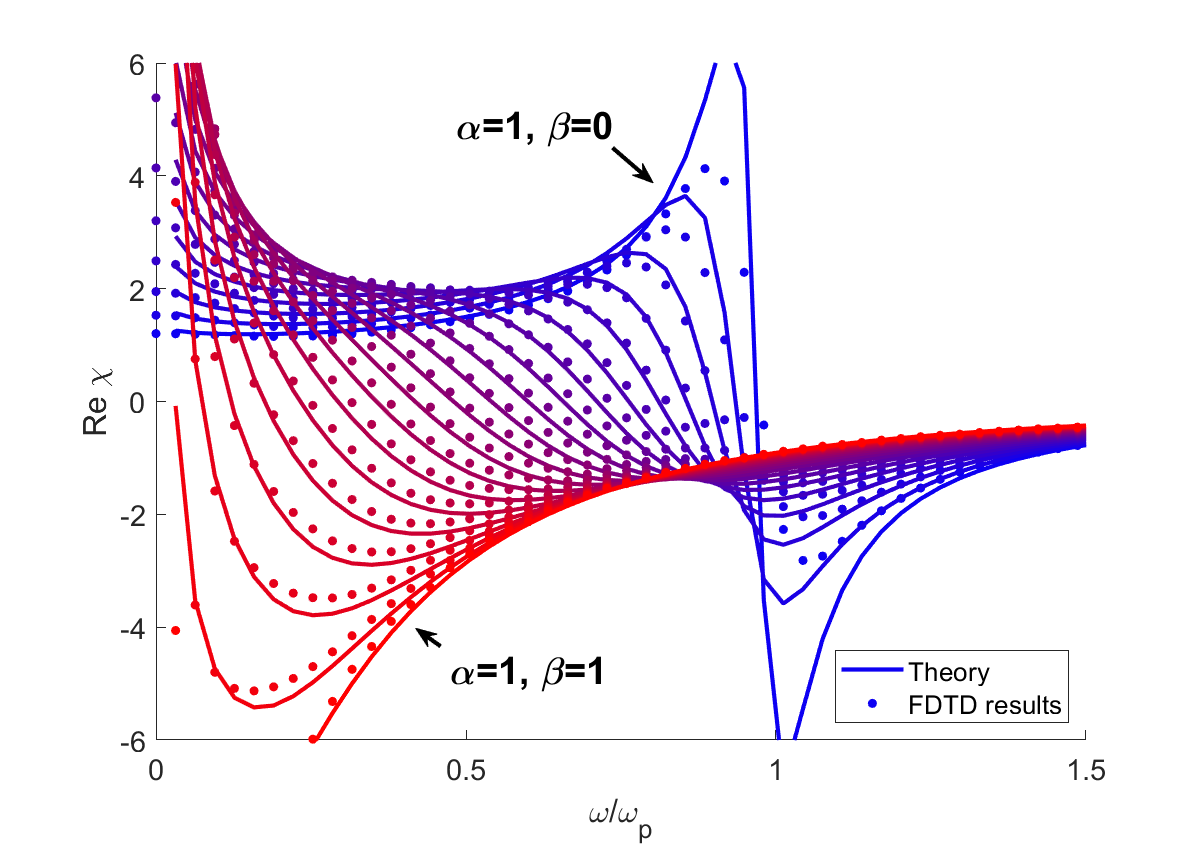}
\caption{Real part of susceptibility calculated analytically from (\ref{koncowe_m}) and numerically for various values of a) $\alpha$ and b) $\beta$.}\label{rys1}
\end{figure*}

Figure \ref{rys1} shows the real part of susceptibility calculated from analytical solution (\ref{koncowe_m}) and obtained in FDTD simulation. 
The medium parameters are set to $\omega_p=0.3$, $\gamma_\beta=0.1$, $\gamma_\alpha=1$. On the Fig. \ref{rys1} a) the value of $\beta$ is set to 1 and $0\leq\alpha\leq 1.5$. Like in a standard Drude model, the susceptibility is negative and for $\alpha\leq 1$, $\chi \rightarrow -\infty$ when $\omega \rightarrow 0$. For higher values of $\alpha$, a distinct minimum of susceptibility occurs. One can see that the accuracy is excellent in the medium frequency range, e. g. $\omega \approx \omega_p$; at the low frequency limit, the large susceptibility corresponds to short wavelength, which approaches the finite spatial step of the simulation. Additionally, waves are highly absorbed in this region, further decreasing prediction accuracy. In the very high frequency regime, accuracy is limited by the fact that the wave period approaches the finite time step. In this region, $\chi \rightarrow 0$ and the value of $\alpha$ has a significant effect on how quickly this asymptote is approached. Fig. \ref{rys1} b) shows the susceptibility calculated for $0\leq\beta\leq 1$ and $\alpha=1$. Here, one can observe a transition from pure Drude model $\beta=1$ to a resonant model with resonance frequency of $\omega=\omega_p$ ($\beta=0$). Again, the calculations become inaccurate in the regions of high absorption at $\omega \rightarrow 0$ and near the resonance. Overall, one can conclude that the model allows for a very significant alteration of the dispersion relation while retaining good stability and accuracy.

\begin{figure}[t!]
\centering
\includegraphics[width=.85\linewidth]{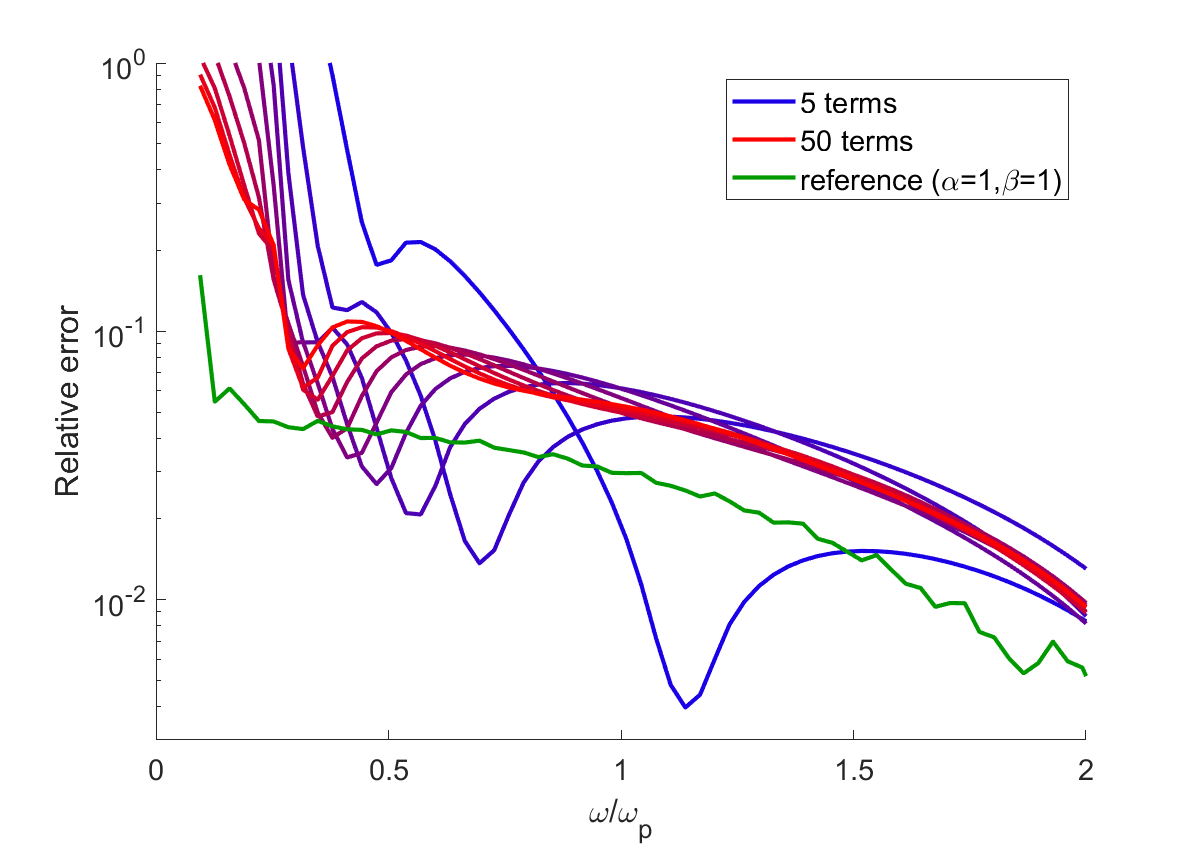}
\caption{Relative error of the model for $\alpha=0.8$, $\beta=0.8$. Standard Drude model (green line) is added for reference.}\label{rys2}
\end{figure}

An important factor for efficient implementation of the proposed model is the number of terms $N$ in the sums in Eq. (\ref{koncowe_num}). The Fig. \ref{rys2} shows the relative error of the numerical susceptibility as a function of frequency, calculated with various numbers of the memory terms $P_{t-k}$. 
One can observe that in general, the error initially quickly decreases with frequency for $\omega < 0.5 \omega_p$, with a further, slower decrease at higher $\omega$. The number of terms has a high impact on the low frequency accuracy. In the solutions with low number of terms, the error has a significant minimum; the calculated values below and above the frequency where the minimum occurs are overestimated and underestimated, accordingly. This can be attributed to the so-called numerical dispersion \cite{Yee}, which introduces small frequency-dependent term to the susceptibility regardless of the medium model. The effect is easily visible on the Fig. \ref{rys1} a) for the case of $\alpha=1.5$; the numerical results (dots) overestimate the susceptibility for $\omega<\omega_p$ and underestimate it beyond that frequency. In majority of the spectral range, the error is greater than in the case of a standard Drude model (Fig. \ref{rys2}, green line) by a factor of 2-3. Apart from the large minimum, the error also exhibits a slight oscillatory behaviour, with amplitude of oscillations reducing with increasing number of terms. One may conclude that the minimum number of summation terms that provide satisfactory accuracy is $N \sim 15$, with relatively little gain from increasing $N$ further. Moreover, the smooth error function with no short-term changes indicates that the model can be easily tuned to maintain a very high accuracy within a chosen, limited spectral range; in the standard FDTD implementation, it is straightforward to add a frequency-independent term $\epsilon_\infty$ to the dielectric susceptibility $\epsilon(\omega)=1+\chi(\omega)$ \cite{Yee}, regardless of the medium model introduced with ADE. By doing so, one can shift the spectral region where the numerical results match the theory exactly. Alternatively, additional Drude-like term can be introduced to ADE to counteract the numerical dispersion. Finally, due to the fact that parameters $\alpha$ and $\beta$ can be changed continuously, one can optimize the accuracy by using slightly different values for theoretical and numerical calculations. The importance of the memory terms and hereditary properties of the medium decreases as the parameters $\alpha$, $\beta$ approach the limit of standard Drude model, e.g. $\alpha=\beta=1$; in such a case, the necessary number of memory terms for accurate computation is reduced. In many applications, only a small modification of the medium is considered \cite{Guia2016}.

In a standard, two-dimensional implementation of FDTD, one has to define three scalar field values at every grid point (for example, two components of magnetic field vector $H_x,H_y$ and one component of electric field $E_z$ perpendicular to the simulation plane).\cite{Yee} The inclusion of Drude or Drude-Lorentz dispersion model as described in \cite{Alsunaidi} adds the medium polarization $P$. Specifically, 1 current ($P_t$) and 1 previous ($P_{t-1}$) value of polarization is needed, resulting in 5 scalar values per grid point. The example implementation of fractional Drude model with 12-term memory adds another 10 scalar values, increasing the memory requirement by a factor of 3. However, it should be noted that the increase is needed only for the part of computational domain that contains the fractional model medium. 

Finally, it should be noted that the coefficients $\alpha_k$, $\beta_k$ in the sums in Eq. (\ref{koncowe_num}) need to be computed once for any given values of $\alpha$, $\beta$ and the calculation of the weighted sum in Eq. (\ref{disc_der}) is essentially a discrete convolution operation, which can be subject to various numerical optimizations. An extensive discussion of numerical application of convolution to calculate a fractional derivative is presented in \cite{Salinas}. 

Another advantage of the presented model is its tunability. By allowing adiabatic changes of $\alpha$ and $\beta$ in the time domain simulation, one achieve a dynamical tuning of the optical properties of the medium.
  
\subsection{Wave propagation in fractional medium}
\begin{figure}[t!]
\centering
\includegraphics[width=.85\linewidth]{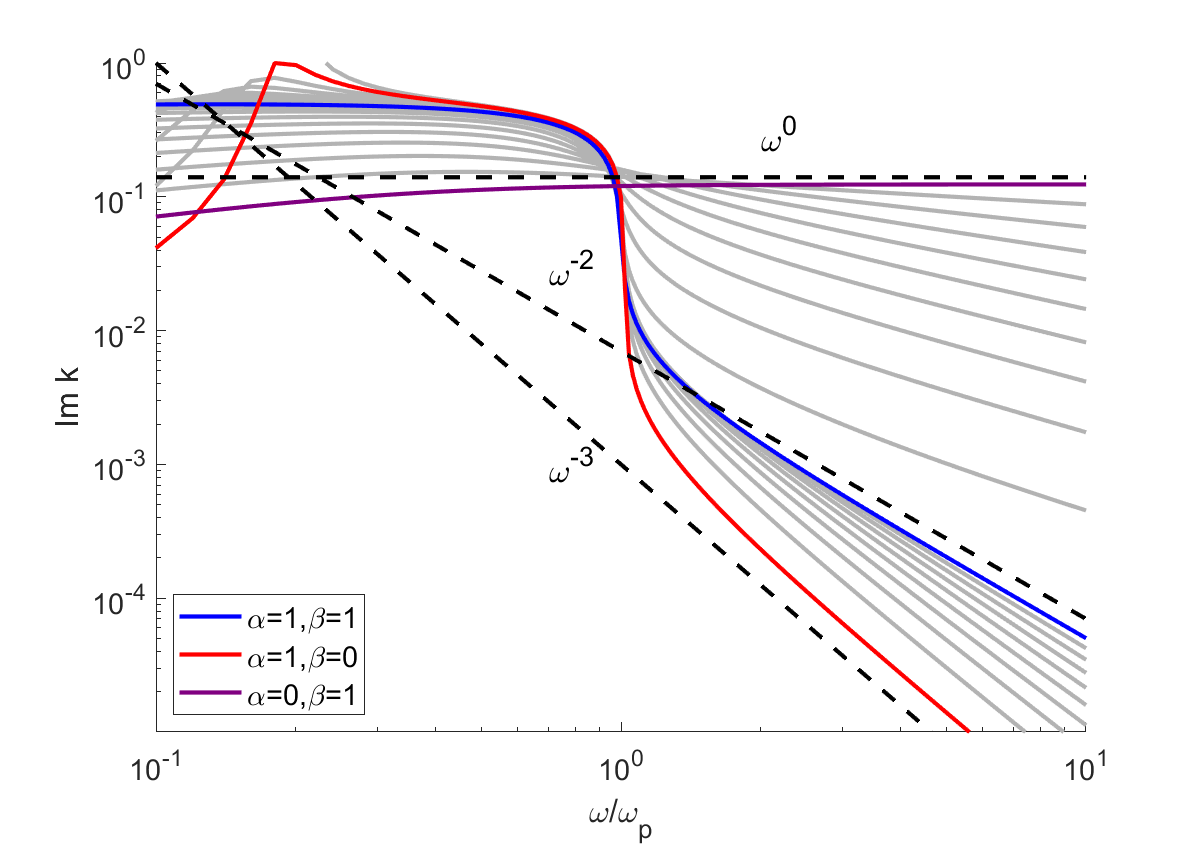}
\caption{Imaginary value of the wave vector as a function of frequency, calculated for various values of $\alpha$ and $\beta$.}\label{rys3}
\end{figure}

Dispersive properties of the medium described by the function $\chi(\omega)$ influence the propagation of electromagnetic waves through the material. It directly affects the permittivity $\epsilon=1+\chi$, refraction index $n=\sqrt{\epsilon}$ and the value of the wave vector $k=\omega n/c$. Assuming a harmonic wave with frequency $\omega$ and wavevector $\vec k$, from the relation (\ref{koncowe_m}) one obtains
\begin{equation}
k^2 = \frac{\omega^2}{c^2}\left(1+\frac{\omega_p^2}{\gamma_\alpha(-i\omega)^{\alpha+1} + \gamma_\beta (-i\omega)^\beta}\right).
\end{equation}
The above relation is nontrivial for real parameters $\alpha$, $\beta$ due to the fact that both terms in denominator introduce separate, frequency-dependent contributions to both real and imaginary part of $\vec k$. One can simplify the problem by assuming that the medium is a small modification of Drude model, with $\alpha=1$ and $\beta \sim 1$; in such a case, in the high frequency limit one obtains
\begin{equation}
k \approx \frac{\omega}{c}\left(1-\frac{\omega_p^2\gamma_\alpha\omega^2}{2\gamma_\alpha^2\omega^4}+i\frac{\omega_p^2\gamma_\beta\omega^\beta}{2\gamma_\alpha^2\omega^4}\right).
\end{equation}
Thus, the imaginary part of $k$ is
\begin{equation}\label{powerlaw}
\mbox{Im}~k \sim \omega^{\beta-3}.
\end{equation}
In a similar manner, one can derive the limit for $\alpha=0$ which is $\mbox{Im}~k \sim \omega^0$. As mentioned in \cite{Holm2011} and \cite{Song2018}, there is a demand on dispersion models where the attenuation (which is proportional to $\mbox{Im}~k$) follows a frequency power law. Fig. \ref{rys3} shows numerically calculated imaginary part of wave vector in the whole spectral range. The discussed limiting cases are shown with color lines, while the transitional spectra with fractional values of $\alpha$, $\beta$ are shown in gray. 
There are two distinct regions $\omega<\omega_p$ and $\omega>\omega_p$; below plasma frequency, the susceptibility is negative and correspondingly $\epsilon \approx 0$, $\mbox{Re}~\vec{k} \approx 0$; the wavevector is thus purely imaginary, which corresponds to highly absorbed, evanescent waves. The parameter $\beta$ has a negligible impact on the absorption in this range. Above the $\omega_p$, the absorption follows a power law (straight line), with the results consistent with Eq. (\ref{powerlaw}). The dependence on the parameter $\alpha$ is more complicated; the steep reduction of absorption at $\omega \approx \omega_p$ for $\alpha=1$ becomes more gradual for smaller values of $\alpha$. In the limit of $\alpha \rightarrow 0$, the absorption becomes almost constant. In contrast to $\beta$, the parameter $\alpha$ has a significant impact on the imaginary part of $k$ in the region $\omega<\omega_p$. While waves with such frequencies are highly attenuated when propagating through the medium, the negative value of permittivity (see Fig. \ref{rys1}) allows for formation of surface plasmons. These collective electron oscillations are highly sensitive to changes of medium properties \cite{my_SPP}, which makes them a particularly promising field of study where the proposed model could be applied. One of the prospective applications of fractional derivative model and resulting power-law dissipation are metallic nanoparticle chains \cite{Song2018}.

\begin{figure}[t!]
\centering
\includegraphics[width=.85\linewidth]{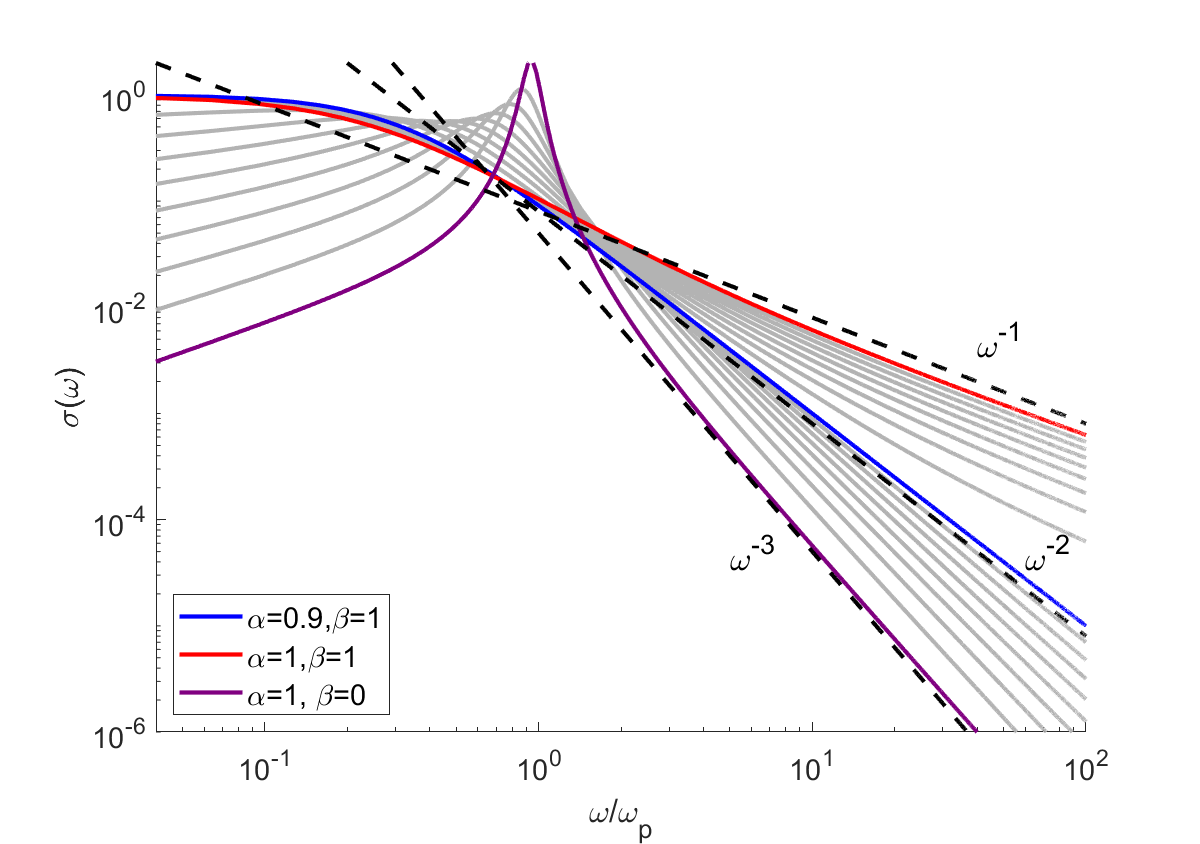}
\caption{The electric conductivity as a function of frequency, calculated for various values of $\alpha$ and $\beta$.}\label{rys4}
\end{figure}

The above mentioned power-law dependence extends to other material functions such as conductivity 
\begin{equation}
\sigma=\epsilon_0\omega\mbox{Im}~\chi
\end{equation}
In the paper \cite{Guia2016} the author proposed a fractional-derivative model based on Eq. (\ref{glowne}), which is transformed to a dimensionless, first-order differential equation describing particle velocity. Our model is consistent with results in \cite{Guia2016} when one sets $\beta=1$ and $\alpha \approx 1$, resulting in a small modification of the standard Drude model. Calculation results for such parameters are shown on the Fig. \ref{rys4}. One can notice that while a  small modification of $\alpha$ has a little impact on the $\sigma$ for $\omega \sim \omega_p$, it dramatically changes the high-frequency behaviour of the medium. As the order of fractional derivative decreases, the slope of the high frequency asymptote increases, in agreement with \cite{Guia2016}. The impact of $\beta$ is much smaller in the high frequency range, but more significant for $\omega<\omega_p$.
\section{Conclusions}
A novel, two-parameter modification of the Drude model based on Grunwald-Letnikov fractional derivative has been presented. The analytical formulas for basic optical functions such as susceptibility and wave vector have been derived and discussed in the context of wave propagation in the medium. A numerical implementation in a FDTD method has been realized, expanding a well-known ADE method and taking advantage of efficient discrete convolution computation. The numerical complexity and accuracy of the approach was discussed. The results indicate that the proposed model is highly flexible and applicable to a wide variety of optical and plasmonic systems, allowing for modelling of other modified Drude models as well as dynamic modulation of medium properties.

\end{document}